\documentclass[twocolumn]{aastex631}
\usepackage{mathrsfs}
\usepackage{graphicx}
\usepackage{subfigure}
\usepackage{epstopdf}
\usepackage{amsmath}
\usepackage{amssymb}
\usepackage{hyperref}
\usepackage{color}

\begin{document}
\title{Constraints on fast radio burst population from the first CHIME/FRB catalog with the Hierarchical Bayesian Inference}

\author{Huan Zhou}
\affiliation{Department of Astronomy, School of Physics and Technology, Wuhan University, Wuhan 430072, China}

\author{Zhengxiang Li}
\affiliation{Department of Astronomy, School of Physics and Astronomy, Beijing Normal University, Beijing 100875, China}
\affiliation{Institute for Frontiers in Astronomy and Astrophysics, Beijing Normal University, Beijing
102206, China}
\email{zxli918@bnu.edu.cn}

\author{Zong-Hong Zhu}
\affiliation{Department of Astronomy, School of Physics and Technology, Wuhan University, Wuhan 430072, China}
\affiliation{Department of Astronomy, School of Physics and Astronomy, Beijing Normal University, Beijing 100875, China}
\email{zhuzh@whu.edu.cn}

\begin{abstract}
Fast Radio Bursts (FRBs) have emerged as one of the most dynamic areas of research in astronomy and cosmology. Despite increasing number of FRBs have been reported, the exact origin of FRBs remains elusive. Investigating the intrinsic redshift distributions of FRBs could provide valuable insights into their possible origins and enhance the power of FRBs as a cosmological probe. In this paper, we propose a hierarchical Bayesian inference approach combining with several viable models to investigate the redshift distribution of the CHIME/FRB catalog 1. By utilizing this method, we aim to uncover the underlying patterns and characteristics of the FRB population, i.e. intrinsic redshift distribution of FRB. Taking uncertainties within the observational data and selection effects into consideration, we obtained that the redshift distribution of FRBs is significantly delayed with respect to that of the star formation history.
\end{abstract}

\section{Introduction}\label{sec1}
Fast Radio Bursts (FRBs) are bright radio transients with durations being a few milliseconds. Since first FRB (FRB 20010724) was discovered in 2007~\citep{Lorimer2007}, hundreds of FRBs have been detected by various radio telescopes, such as the Australian Square Kilometer Array Pathfinder (ASKAP), the Canadian Hydrogen Intensity Mapping Experiment (CHIME), the Five-hundred-meter Aperture Spherical radio telescope (FAST), and Deep Synoptic Array-110 (DSA-110), etc. While their radiation mechanism and progenitors remains unknown, several theories suggest they could be linked to exotic astrophysical phenomena such as magnetars, black holes or neutron star mergers~\citep{Petroff2019,Cordes2019,Zhang2020,Xiao2021}. However, several distinctive and valuable observational properties of FRBs, such as their short durations, cosmological origins, and high all-sky event rate, have been identified as potential probes for cosmology and astrophysics~\citep{Gao2014,Wei2015, Munoz2016,Li2018,Li2020,Liao2020,Zhou2022}.

The FRB population, i.e. the energy and the redshift distribution, has been extensively studied in many literatures~\citep{Zhang2019,Pleunis2021, Zhang2021, James2022a, James2022b, Qiang2022, Zhang2022, Shin2022,Zhang2023, Chen2024, Lin2024a, Lin2024b,Gupta2025}. Based on assumptions for dispersion measures (DMs) from host galaxies and the Milky Way halo, it was found that, independent of the redshift distribution, the intrinsic energy distribution of FRBs may follow a power law distribution with a high-energy exponential cutoff~\citep{Luo2020,Lu2020}. However, due to different predictions of progenitor theories, the redshift distribution is not as clear as the energy distribution. For instance, a delayed FRB redshift distribution compared to the star formation history (SFH) was obtained from the FRB data detected by the Parkes and ASKAP~\citep{Zhang2021}. Meanwhile, several independent studies also rule out the hypothesis that FRBs trace the SFH with the CHIME/FRB catalog 1, but the exact FRB population remains unconstrained~\citep{Zhang2019,Qiang2022, Zhang2022, Chen2024,Lin2024a, Lin2024b,Gupta2025}. Above all, the intrinsic redshift distribution of FRBs remains a topic of extensive debate.

In this paper, compared with the Kolmogorov-Smirnov (KS) test method~\citep{Qiang2022, Zhang2022} and Bayesian method with the cumulative distribution function~\citep{Lin2024a, Lin2024b}, we propose the hierarchical Bayesian inference (HBI) method to investigate the characteristics of the FRB population on the basis of the well-measured FRB catalog 1 reported by CHIME~\citep{CHIME2021}. This paper is organized as follows: In Section~\ref{sec2}, we introduce the FRB observations and the hierarchical Bayesian inference used to derive constraints on the population hyperparameters of FRB. In Section~\ref{sec3}, we apply this approach to the selected FRB data and present the corresponding results. Finally, Section~\ref{sec4} presents conclusions and discussions. Throughout this paper, we use the concordance $\Lambda$CDM cosmology with the best-fitting parameters from the latest $Planck$ cosmic microwave background (CMB) observations~\citep{Planck2018}.

\section{Methodology}\label{sec2}

\begin{figure}
    \centering
     \includegraphics[width=0.45\textwidth, height=0.32\textwidth]{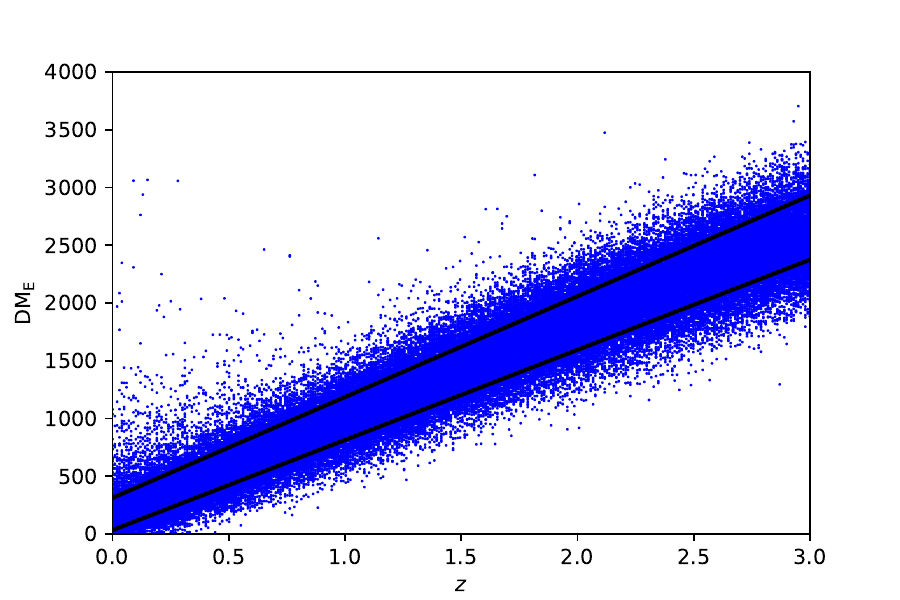}
     \caption{\textbf{The blue scatter represents the ${\rm DM}_{\rm E}-z$ relation.} The upper and lower black lines represent $\sigma_+(z)$ and $\sigma_-(z)$ in  $1\sigma$ confidence region, respectively. }\label{fig0}
\end{figure}

\subsection{Fast Radio Burst Population Models}\label{sec2-1}
The redshift distribution of Fast Radio Bursts (FRBs) can be inferred using the dispersion measure (DM) as a proxy. DM is the integral of the free electron number density along the line of sight, and it can be directly obtained from the dynamical spectra of the burst. The distance and redshift of a detected FRB can be approximately estimated from its observed DM, which is proportional to the number density of free electron along the line of sight and is usually decomposed into the following three ingredients~\citep{Deng2014,Macquart2020},
\begin{equation}\label{eq2-1}
{\rm DM_{\mathrm{obs}}}={\rm DM_{MW}}+{\rm DM_{IGM}}+\frac{\rm DM_{host}}{1+z}.
\end{equation}
${\rm DM_{MW}}$ is the contribution from the Milky Way as 
\begin{equation}\label{eq2-2}
{\rm DM_{MW}}={\rm DM_{MW,ISM}}+{\rm DM_{MW,halo}},
\end{equation}
where ${\rm DM_{MW,ISM}}$ interstellar medium, which can be estimated using the Milky Way electron density models, such as the NE2001 model~\citep{Cordes2002} and YMW16 model~\footnote{We have tested the influence of these two models on our results and found that the impact is negligible. Therefore, our following analysis are based on the NE2001 model.}~\citep{Yao2017}. 
Previous studies suggested that sources with low DM imply small contributions, $\sim25~\rm{pc~cm^{-3}}$, from the halos of M81 and the Milky Way~\citep{Bhardwaj2021, Cordes2021}. Therefore, ${\rm DM_{MW,halo}}$ is approximately estimated to be $25~\rm{pc~cm^{-3}}$ in this paper. ${\rm DM_{IGM}}$ represents DM contribution from intergalactic medium (IGM). In the standard $\Lambda \rm{CDM}$ model the average value of the ${\rm DM_{IGM}}$ can be calculated as~\citep{Deng2014}
\begin{equation}\label{eq2-3}
\begin{split}
\langle{\rm DM_{IGM}(z)}\rangle=\frac{3H_0\Omega_{\rm b}f_{\rm IGM}f_{\rm e}}{8\pi m_{\rm p}}\times\\
\int_0^z\frac{1+z}{\sqrt{\Omega_{\rm m}(1+z)^3+\Omega_{\Lambda}}}dz,
\end{split}
\end{equation}
where $H_0$ is the Hubble constant, $\Omega_{\rm m}$ is the present-day matter density, $m_{\rm p}$ is the proton mass, $f_{\rm IGM} = 0.84$ is the baryon mass fraction in IGM, and $f_{\rm e} = 7/8$ is the electron fraction. Following previous works~\citep{Qiang2020}, we consider the $\sigma_{\rm IGM}$ was given by a simple power-law function as
\begin{equation}\label{eq2-4}
\sigma_{\rm IGM}\approx 173.8z^{0.4}~{\rm pc~cm^{-3}}.
\end{equation}
In addition, ${\rm DM_{host}}$ represents the contribution from host galaxy, and it is usually modeled as a log-normal distribution~\citep{Macquart2020}
\begin{equation}\label{eq2-5}
\begin{split}
p({\rm DM_{host}}|\mu_{\rm h}, \sigma_{\rm h})=\frac{1}{\sqrt{2\pi}{\rm DM_{host}}\sigma_{\rm h}}\times\\
\exp\bigg[-\frac{(\ln({\rm DM_{host}})-\mu_{\rm h})^2}{2\sigma_{\rm h}^2}\bigg]
\end{split}
\end{equation}
where the parameters $\mu_{\rm h}$ and $\sigma_{\rm h}$ are the mean and standard deviation of $\ln({\rm DM_{host}})$, and Tang et al. (2023) has given the well constrains ($\mu_{\rm h}$, $\sigma_{\rm h}$) from 17 well-localized FRBs~\citep{Tang2023}. Thus, subtracting the known ${\rm DM_{MW}}$ from $\rm DM_{\rm obs}$ in Equation~(\ref{eq2-1}) ,it is convenient to introduce the extragalactic ${\rm DM}_{\rm E}$ of an FRB as the observed quantity
\begin{equation}\label{eq2-5f}
{\rm DM}_{\rm E}\equiv{\rm DM_{IGM}}+\frac{\rm DM_{host}}{1+z}.
\end{equation}
As shown in Figure~\ref{fig0}, the blue scatter represents the ${\rm DM}_{\rm E}-z$ relation which is inferred from the contributions of ${\rm DM_{\rm IGM}}$ and ${\rm DM_{\rm host}}$ with their uncertainties. In addition, $\sigma_+(z)$ (the upper black line in Figure~\ref{fig0}) and $\sigma_-(z)$ (the lower black line in Figure~\ref{fig0}) represent the best polynomial fit of the boundary of  $1\sigma$ confidence region the for ${\rm DM}_{\rm E}-z$ relation, respectively.

The observed FRB redshift rate distribution is from the intrinsic event rate density of FRB $dN/(dtdV)\equiv\mathcal{R}(z|\Phi)$, the most well-motivated model is the SFH model. Based on the gamma-ray burst population does not track SFH and the enhanced evolution in the GRB rate can be parameterized as $\mathcal{R}(z)\propto \varepsilon(z){\rm SFH}(z)$~\citep{Kistler2007,Yukse2008}, we adopt three kinds of best-fit SFH-related models discussed in previous studies~\citep{Qiang2022,Lin2024a}. 
\begin{itemize}
\item\textbf{1.Two-segment (TSE) model:}
\begin{equation}\label{eq2-6}
\begin{split}
\mathcal{R}(z|\gamma_1,\gamma_2,z_{\rm t})=R_{{\rm TSE},0}\times\\
\frac{(1+z)^{\gamma_1}}{1+((1+z)/(1+z_{\rm t}))^{\gamma_1+\gamma_2}}{\rm SFH}(z),
\end{split}
\end{equation}
where $R_{{\rm TSE},0}$ is the local event rate in ${\rm Gpc^{-3}}yr^{-1}$, and ${\rm SFH}(z)$ is the empirical SFH model~\citep{Madau2017} as
\begin{equation}\label{eq2-7}
{\rm SFH}(z)=\frac{(1+z)^{2.6}}{1+((1+z)/3.2)^{6.2}},
\end{equation}
which is expected to behave like $(1+z)^{2.6}$ and $(1+z)^{-3.6}$ at low and high redshifts respectively.

\item\textbf{2.Power law with an exponential cutoff (CPL) model:}
\begin{equation}\label{eq2-8}
\begin{split}
\mathcal{R}(z|\gamma,z_{\rm c})=R_{{\rm CPL},0}\times\\
(1+z)^{\gamma}\exp(-z/z_{\rm c}){\rm SFH}(z),
\end{split}
\end{equation}
which is expected to behave like $(1+z)^{2.6+\gamma}$ and exponential cutoff at low and high redshifts respectively.

\item\textbf{3.Non-SFH-based two-segment redshift distribution (TSRD) model:}
\begin{equation}\label{eq2-9}
\begin{split}
\mathcal{R}(z|\lambda,\kappa,z_{\rm p})=R_{{\rm TSRD},0}\times\\
\frac{(1+z)^{\lambda}}{1+((1+z)/(1+z_{\rm p}))^{\lambda+\kappa}},
\end{split}
\end{equation}
if $z_{\rm p}$ is much larger than 0 and $\lambda+\kappa$ is positive, this model is expected to behave like $(1+z)^{\lambda}$ and $(1+z)^{-\kappa}$ at low and high redshifts respectively.
\end{itemize}
Therefore, the population hyperparameters of three models $\Phi$ are
\begin{equation}\label{eq3-10}
\Phi=[\gamma_1,\gamma_2,z_{\rm t}, \gamma,z_{\rm c},\lambda,\kappa,z_{\rm p}].
\end{equation}
Consistent with previous works, we set similar prior distributions for all the hyperparameters $\Phi$ as shown in Table~\ref{tab1}.

{\renewcommand{\arraystretch}{1.5}
\begin{table}[!ht]
\centering
\setlength{\tabcolsep}{4.5mm}{\begin{tabular}{ccc}
\hline
Model & Hyperparameter $\Phi$ & Prior\\
\hline
{} & $\gamma_1$  & $\mathcal{U}[-10,~10]$\\ 
{TSE} & $\gamma_2$ & $\mathcal{U}[-10,~10]$\\ 
{} & $z_{\rm t}$ & $\mathcal{U}[0,10]$\\ 
\hline
{CPL} & $\gamma$  & $\mathcal{U}[-10,~10]$\\ 
{} & $z_{\rm c}$ & $\mathcal{U}[0,~10]$\\ 
\hline 
{} & $\lambda$ & $\mathcal{U}[-10,~10]$\\ 
{TSRD} & $\kappa$ & $\mathcal{U}[-10,10]$\\ 
{} & $z_{\rm p}$  & $\mathcal{U}[0,~10]$\\
\hline
\end{tabular}}
\caption{\label{tab1} Population hyperarameters $\Phi=[\gamma_1,\gamma_2,z_{\rm t}, \gamma,z_{\rm c},\lambda,\kappa,z_{\rm p}]$ and their prior distributions used in the HBI.}
\end{table}}

\subsection{Hierarchical Bayesian Inference}\label{sec2-2}
For population hyperparameters of FRB $\Phi$, and $N_{\rm obs}$ detections of FRB events $d = [d_1,...d_{N_{\rm obs}}]$~\footnote{In later analysis, $d_i$ represents the DM for single observed non-repeating FRB event in CHIME catalog 1.}, the likelihood follows a Poisson distribution without considering measurement uncertainty and selection effect is
\begin{equation}\label{eq3-1}
p(d|\Phi)\propto N(\Phi)^{N_{\rm obs}}e^{-N(\Phi)}.
\end{equation}
However, with measurement uncertainty and selection effect taken into account, the likelihood for $N_{\rm obs}$ FRB observations can be characterized by the inhomogeneous Poisson process as~\citep{Mandel2019, LVK2022, Mastrogiovanni2024}
\begin{equation}\label{eq3-2}
\begin{split}
&p(d|\Phi)\propto N(\Phi)^{N_{\rm obs}}e^{-N(\Phi)\xi(\Phi)}\times\\
&\prod_{i}^{N_{\rm obs}}\int dz L(d_i|z)p_{\rm pop}(z|\Phi),
\end{split}
\end{equation}
where $L(d_i|z)$ is the likelihood of one FRB. We assume the likelihood $L(d_i|z)$ for data follows the Gaussian distribution with the error coming from ${\rm DM_{IGM}}$ and ${\rm DM_{host}}$ as
\begin{equation}\label{eq3-2f}
L(d_i|z)=\frac{1}{\sqrt{2\pi}\sigma_{z,i}}\exp\bigg[-\frac{(z-\bar{z}_i)^2}{2\sigma_{z,i}^2}\bigg],
\end{equation}
where $\bar{z}_i$ and $\sigma_{z,i}$ come from the ${\rm DM}_{{\rm E}}-z$ relation of each FRB as
\begin{equation}\label{eq3-2ff}
\begin{split}
\bar{z}=\frac{\sigma^{-1}_{+}({\rm DM}_{{\rm E}})+\sigma^{-1}_{-}({\rm DM}_{{\rm E}})}{2}~~\\
\sigma_{z}=\frac{\sigma^{-1}_{+}({\rm DM}_{{\rm E}})-\sigma^{-1}_{-}({\rm DM}_{{\rm E}})}{2},
\end{split}
\end{equation}
where $\sigma^{-1}_{+}({\rm DM}_{{\rm E}})$ and $\sigma^{-1}_{-}({\rm DM}_{{\rm E}})$ represent the inverse function of  the best polynomial fit for $\sigma_{+}(z)$ and $\sigma_{-}(z)$ shown in Figure~\ref{fig0}, respectively. In Equation~(\ref{eq3-2}), $N(\Phi)$ is the total number of FRBs in the model characterized by the set of population parameters $\Phi$ as
\begin{equation}\label{eq3-3}
N(\Phi)=\int dz T_{\rm obs}\mathcal{R}(\lambda|\Phi)\frac{1}{1+z}\frac{dV_{\rm c}}{dz},
\end{equation}
where $dV_{\rm c}/{dz}$ is the differential comoving volume, the factor $1/(1 + z)$ accounts for the cosmological time dilation from the source frame to the detector frame, and we define $T_{\rm obs}$ as the effective observing time which is $100\%$ operational time of CHIME~\footnote{CHIME operates nominally 24 hr per day during 25 July 2018 and 1 July 2019, but CHIME was not fully operational $100\%$ of the time as some reasons shown in Section 2.2~\citep{CHIME2021}}. In addition, $p_{\rm pop}(z|\Phi)$ is the normalized distribution of redshifts in FRBs as
\begin{equation}\label{eq3-4}
p_{\rm pop}(z|\Phi)=\frac{1}{N(\Phi)}
\bigg[T_{\rm obs}\mathcal{R}(z|\Phi)
\frac{1}{1+z}\frac{dV_{\rm c}}{dz}\bigg].
\end{equation}
Meanwhile, the detected fraction $\xi(\Phi)$ of FRBs observed by the CHIME telescope can be defined as 
\begin{equation}\label{eq3-5}
\xi(\Phi)\equiv\int dz P_{\rm det}(z)p_{\rm pop}(z|\Phi),
\end{equation}
where $P_{\rm det}(z)$ is the detection probability that depends on the source and instrument. The selection effect of the CHIME telescope is determined by detection efficiency as a function $\eta_{\rm det}$ of the specific fluence~\citep{Zhang2022,Qiang2022,Lin2024a,Lin2024b}
\begin{equation}\label{eq3-6}
\eta_{\rm det}(F_{\nu})=\bigg(\frac{\log(F_{\nu})-\log(F_{\nu,\min})}{\log(F_{\nu,\max})-\log(F_{\nu,\min})}\bigg)^n.
\end{equation}
This model assumes a `grey zone’ between the minimum specific threshold fluence $F_{\nu,\min}$, and the maximum specific threshold fluence $F_{\nu,\max}$. FRBs with $F_{\nu}<F_{\nu,\min}$, are undetectable
hence $\eta_{\rm det} = 0$, while above $F_{\nu,\max}$ all FRBs are detectable hence $\eta_{\rm det} = 1$. 
Assuming a flat radio spectrum, the observed specific fluence $F_{\nu}$ for a mock FRB with redshift $z$ and 
isotropic energy $E$ is~\citep{Zhang2018,James2022a,Zhang2021}
\begin{equation}\label{eq3-7}
F_{\nu}=\frac{(1+z)^{2+\beta}}{4\pi d_{\rm L}^2\Delta\nu}E,
\end{equation}
where $\beta$ is the spectrum index $F_{\nu}\propto \nu^{\beta}$, $d_{\rm L}$ is the luminosity distance, and $\Delta\nu$ is the bandwidth in which the FRB is detected.~\citet{Macquart2019} have fitted $\beta$ with 23 FRBs detected by ASKAP in Fly’s Eye mode, finding best-fit value $\beta=-1.5$. Like~\citet{Luo2020,Lu2020}, we assume the model of energy function $p(E)$ as power-law with high-energy exponential cutoff
\begin{equation}\label{eq3-8}
p(E)=\Phi_0\bigg(\frac{E}{E_{\rm c}}\bigg)^{-\alpha}\exp\bigg(-\frac{E}{E_{\rm c}}\bigg),
\end{equation}
where $\alpha$ is the \textbf{power-law index}, $E_{\rm c}$ is the break energy, and $\Phi_0$ is a normalization constant. Previous works showed
that $\alpha\sim1.8$ is consistent with observations~\citep{Luo2020,Lu2020}, and~\citet{Luo2020} suggested an energy cut-off at  $E_{\rm c}\sim3\times 10^{41} {\rm erg}$. By marginalizing the local event rate ($R_{{\rm TSE},0},~R_{{\rm CPL},0},~R_{{\rm TRSD},0}$) with a log-uniform prior for it, the posterior distribution $p(\Phi|d)$ could be obtained
\begin{equation}\label{eq3-9}
\begin{split}
p(\Phi|d)=\frac{p(d|\Phi)p(\Phi)}{Z_{\mathcal{M}}}\\
\propto \frac{p(\Phi)}{Z_{\mathcal{M}}}\xi(\Phi)^{-N_{\rm obs}}&\prod_{i}^{N_{\rm obs}}\int dz L(d_i|z)p_{\rm pop}(z|\Phi),
\end{split}
\end{equation}
where $Z_{\mathcal{M}}$ is the Bayesian evidence for the population model $\mathcal{M}$, $p(\Phi)$ is prior distribution for the hyperparameters $\Phi$.

\begin{figure*}
    \centering
     \includegraphics[width=0.32\textwidth, height=0.32\textwidth]{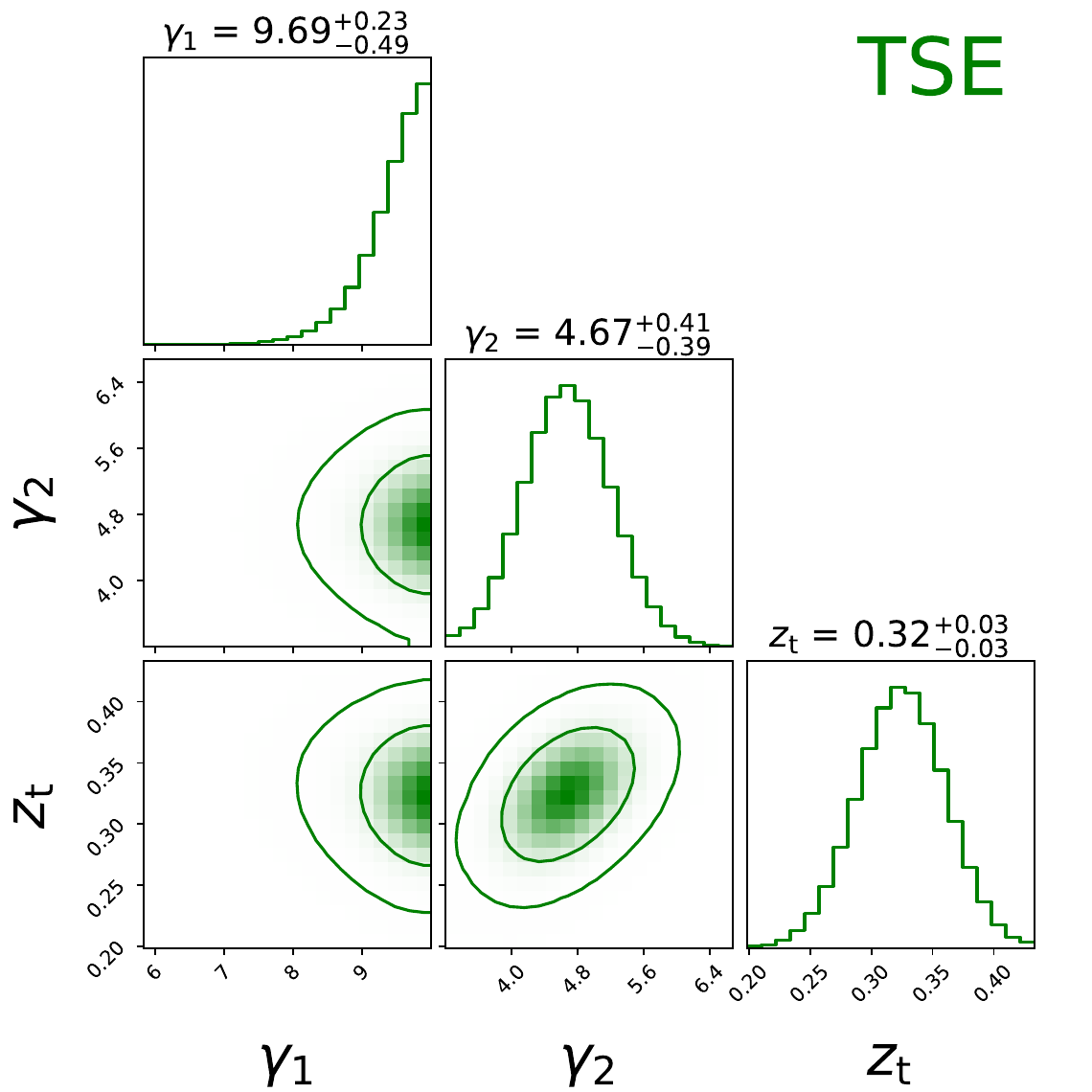}
     \includegraphics[width=0.32\textwidth, height=0.32\textwidth]{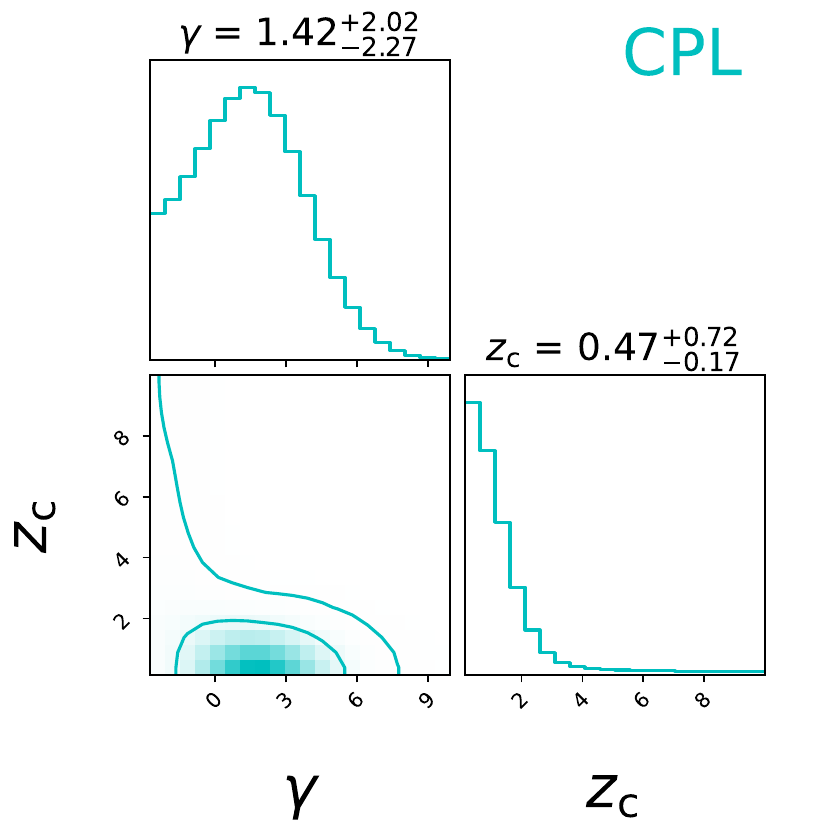}
    \includegraphics[width=0.32\textwidth, height=0.32\textwidth]{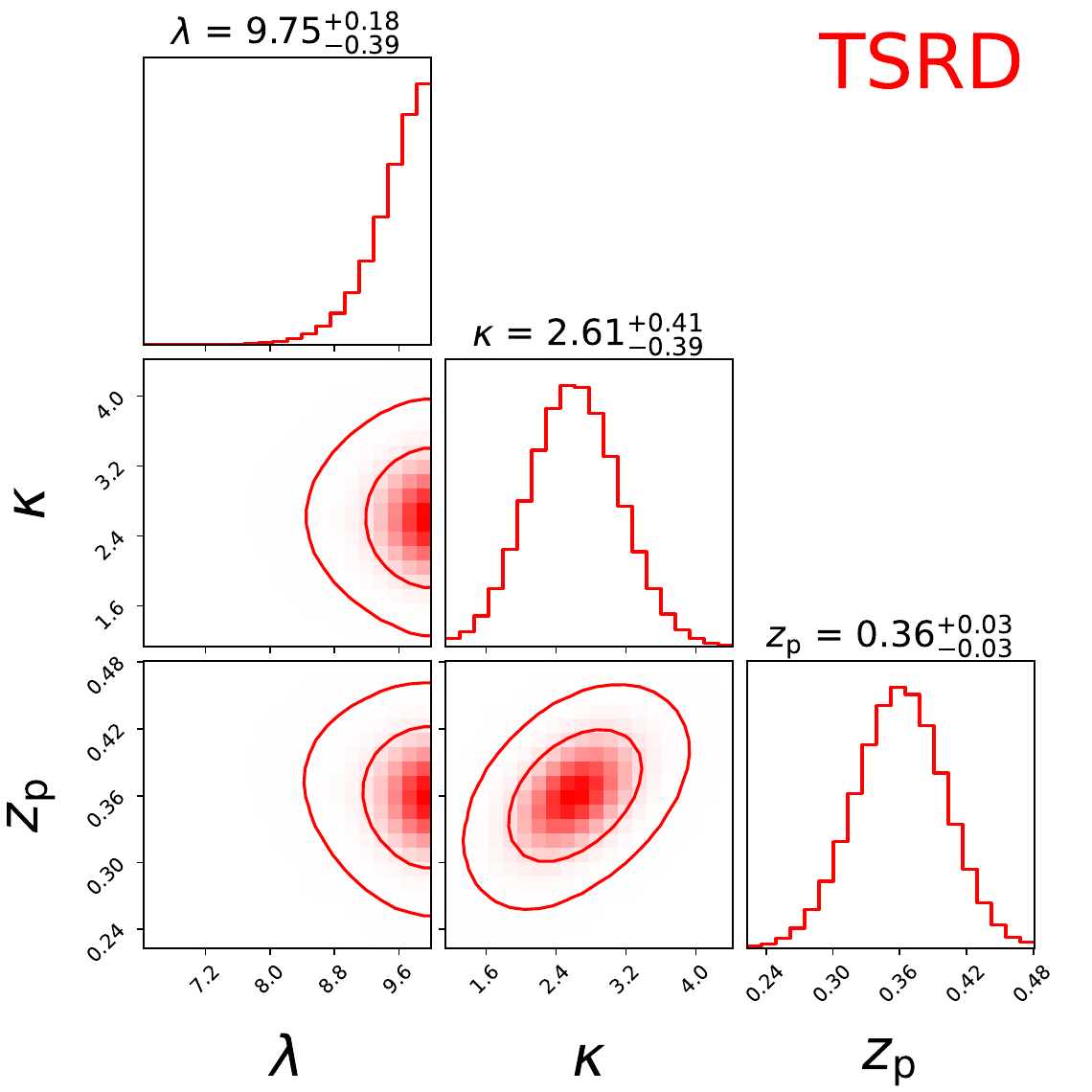}
     \caption{The contour plot with $2~\sigma$ uncertainty for the TSE model (left), CPL model (middle) and TSRD model (right).}\label{fig1}
\end{figure*}

\section{Results}\label{sec3}
In this section, we use the first CHIME/FRB catalog with well measured ${\rm DM_{\rm obs}}$ to constrain FRB population hyperparameters, which are detected from 2018 July 25 to 2019 July 1~\footnote{https://www.chime-frb.ca/catalog}~\citep{CHIME2021}. The first
CHIME/FRB catalog contains 536 bursts in total, i.e. 474 non-repeating bursts and 62 repeating bursts corresponding to 18 different FRB sources. Similar to previous works~\citep{Zhang2022,Qiang2022,Lin2024a,Lin2024b}, here we only select the FRBs with ${\rm DM}_{\rm E}\geq200~{\rm pc~cm^{-3}}$ (in total 415 FRBs).

\begin{figure}
    \centering
     \includegraphics[width=0.45\textwidth, height=0.32\textwidth]{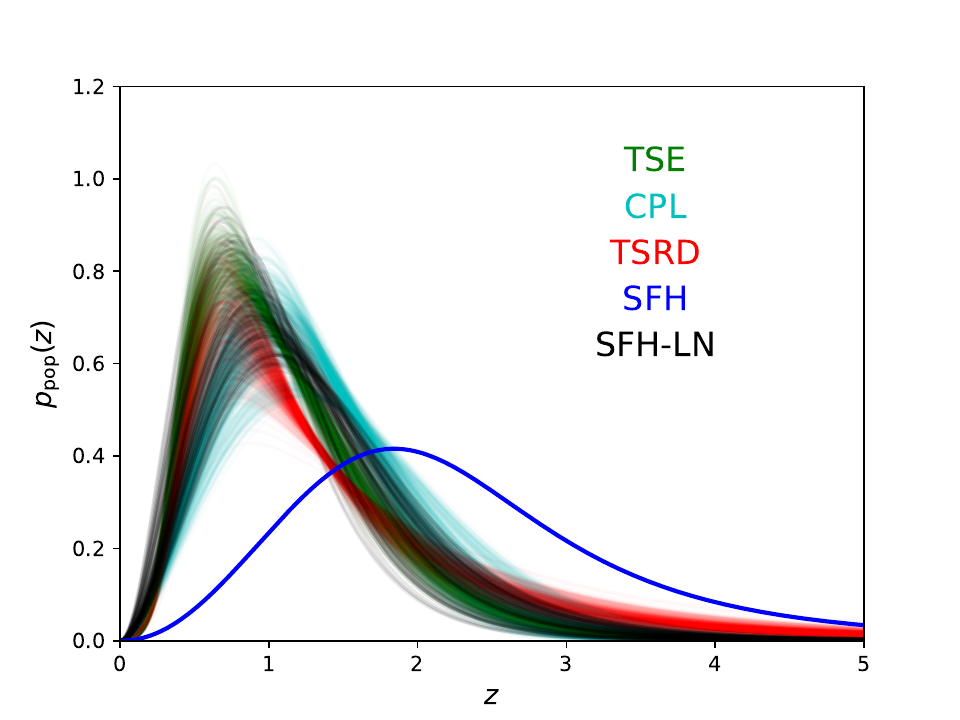}
     \caption{The redshift distribution of FRBs derived from the reconstruction procedure with the posterior distributions for TSE, CPL and TSRD models respectively. For comparison, we also give the redshift distribution from SFH model. }\label{fig2}
\end{figure}

Then we incorporate the above-mentioned FRBs sample into the \textbf{EMCEE}~\citep{emcee} with the posterior Equation~(\ref{eq3-9}) to estimate population hyperparameters ($\Phi=[\gamma_1,\gamma_2,z_{\rm t}, \gamma,z_{\rm c},\lambda,\kappa,z_{\rm p}]$) from three models. Our results are summarized in Figure~\ref{fig1} and Table~\ref{tab2}. We find that the inferred FRB redshift distribution is tilted lower away from the predicted SFH peak, which is consistent with~\citet{Lin2024a,Lin2024b} works. To quantify the goodness of three models, we perform model comparison statistics by using the Bayesian Information Criterion (BIC)~\citep{Schwarz1978}, which is expressed as
\begin{equation}\label{eq4-1}
{\rm BIC}_{\mathcal{M}}=-2\ln(\mathcal{L}_{\max,\mathcal{M}})+k_{\mathcal{M}}\ln(N_{\rm obs}),
\end{equation}
where $k_{\mathcal{M}}$ represent the total number of population hyperparameters, and  $\mathcal{L}_{\max,\mathcal{M}}$ is the maximum likelihood $p(d|\Phi)$ value for the $N_{\rm obs}$ FRB events in the framework of $\mathcal{M}$ model. $\Delta {\rm BIC}\geq5$ or $\geq10$ indicates that there is “strong” or “decisive” evidence against the considered model with respect to the
fiducial model. With the results listed in Table~\ref{tab2}, we have the rank of fits as ${\rm TSRD}\gtrsim{\rm TSE}\gg{\rm CPL}$. However, we find that the constrains of some hyperparameters in the three models, i.e. $\gamma_1$, $\gamma$, and $\lambda$, tend to be larger in range, which indicates that the redshift distribution of FRBs is steeper in the low redshift interval. 

In order to compare the SFH model and its time delay model, the redshift distributions of FRBs derived from the reconstruction with the posterior distributions for three kind of models have shown in Figure~\ref{fig2}. In Figure~\ref{fig2}, SFH-LN represents the redshift distribution corresponding to the SFH model with the log-normal time delay distribution $f(t|\tau_0,\sigma_0)$ as 
\begin{equation}\label{eq4-2}
\mathcal{R}(z)\propto\int_{0}^{t_{\max}}{\rm SFH}(t(z)-t_{\rm d})f(t_{\rm d}|\tau_0,\sigma_0) dt_{\rm d}.
\end{equation}
For the sake of comparison, we also take the parameters $\tau_0$ and $\sigma_0$ as uniform distributions $\mathcal{U}[5,~10]$ Gyr and $\mathcal{U}[0.8,~1.5]$, respectively. We find that the distributions of FRBs from our results do not trace the history of star formation, and  there is the tendency for the redshift distribution of FRBs to favor SFH model with a time delay. Moreover, this is consistent with conclusions from~\citet{Zhang2022,Chen2024,Lin2024b,Gupta2025} works.

{\renewcommand{\arraystretch}{1.5}
\begin{table}[!ht]
\centering
\setlength{\tabcolsep}{4.5mm}{\begin{tabular}{ccc}
\hline
Model & Hyperparameter $\Phi$ & $\Delta {\rm BIC}$\\
\hline
{} & $\gamma_1=9.69^{+0.23}_{-0.49}$  & {}\\ 
{TSE} & $\gamma_2=4.67^{+0.41}_{-0.39}$ & {0}\\ 
{} & $z_{\rm t}=0.32^{+0.03}_{-0.03}$ & {}\\ 
\hline
{CPL} & $\gamma=1.42^{+2.02}_{-2.27}$  & {98.87}\\ 
{} & $z_{\rm c}=0.47^{+0.72}_{-0.17}$ & {}\\ 
\hline
{} & $\lambda=9.75^{+0.18}_{-0.39}$ & {}\\ 
{TSRD} & $\kappa=2.61^{+0.41}_{-0.39}$ & {-2.10}\\ 
{} & $z_{\rm p}=0.36^{+0.03}_{-0.03}$  & {}\\ 
\hline
\end{tabular}}
\caption{\label{tab2} The optimal Population hyperarameters $\Phi=[\gamma_1,\gamma_2,z_{\rm t}, \gamma,z_{\rm c},\lambda,\kappa,z_{\rm p}]$ and their $1~\sigma$ uncertainties. For comparing the TSE population model with other models, we also list $\Delta{\rm BIC}$.}
\end{table}}

\section{Conclusions and Discussion}\label{sec4}
Nowadays, FRBs have become one of the most significant probes for astrophysical and cosmological purposes. However, the origin of FRBs remains unknown. Studies of the intrinsic FRB distributions may provide insights into their possible origins. In this paper, based on FRB data from the first CHIME/FRB catalog, we propose the HBI method to constrain the FRB population information, i.e. intrinsic redshift distribution of FRB. We find the current FRB sample does not trace the history of star formation. This conclusion is consistent with the findings of many previous studies~\citep{Zhang2019,Zhang2021, Qiang2022, Zhang2022, Zhang2023, Chen2024, Lin2024a, Lin2024b,Gupta2025}, which is also understandable, because the current inferred redshift distribution of is mostly concentrated below 2.

With the rapid increase of the number of FRBs detected by poweful wide-field surveys (like CHIME, SKA, and DSA-110) and more stringent restrictions on the FRB host environment in the future, the population characteristics of FRBs will be increasingly restricted, which will help us to study their origins. In addition, similar to using compact object populations and gravitational wave events in GWTC-3 to limit cosmological models~\citep{LVK2022, Mastrogiovanni2024}, since $\rm DM_{\rm E}$ of FRB contains cosmological information, it is foreseen that FRB can also be take as a ``dark siren" for constraining cosmology.

\section{Acknowledgements}
We are grateful to Dachun Qiang for helpful discussion. This work is supported by the China National Postdoctoral Program for Innovative Talents under Grant No.BX20230271; National Key Research and Development Program under Grant No.2024YFC2207400; National Key Research and Development Program of China Grant No. 2021YFC2203001; National Natural Science Foundation of China under Grants Nos.11920101003, 12021003, 11633001, 12322301, and 12275021; the Strategic Priority Research Program of the Chinese Academy of Sciences, Grant Nos. XDB2300000 and the Interdiscipline Research Funds of Beijing Normal University.

\bibliography{ref}{}

\begin{thebibliography}{}
\expandafter\ifx\csname natexlab\endcsname\relax\def\natexlab#1{#1}\fi
\providecommand{\url}[1]{\href{#1}{#1}}
\providecommand{\dodoi}[1]{doi:~\href{http://doi.org/#1}{\nolinkurl{#1}}}
\providecommand{\doeprint}[1]{\href{http://ascl.net/#1}{\nolinkurl{http://ascl.net/#1}}}
\providecommand{\doarXiv}[1]{\href{https://arxiv.org/abs/#1}{\nolinkurl{https://arxiv.org/abs/#1}}}

\bibitem[{Abbott {et~al.}(2023)}]{LVK2022}
Abbott, R., {et~al.} 2023, Phys. Rev. X, 13, 011048,
  \dodoi{10.1103/PhysRevX.13.011048}

\bibitem[{Aghanim {et~al.}(2020)}]{Planck2018}
Aghanim, N., {et~al.} 2020, Astron. Astrophys., 641, A6,
  \dodoi{10.1051/0004-6361/201833910}

\bibitem[{Amiri {et~al.}(2021)}]{CHIME2021}
Amiri, M., {et~al.} 2021, Astrophys. J. Supp., 257, 59,
  \dodoi{10.3847/1538-4365/ac33ab}

\bibitem[{{Bhardwaj} {et~al.}(2021){Bhardwaj}, {Gaensler}, {Kaspi},
  {Landecker}, {Mckinven}, {Michilli}, {Pleunis}, {Tendulkar}, {Andersen},
  {Boyle}, {Cassanelli}, {Chawla}, {Cook}, {Dobbs}, {Fonseca}, {Kaczmarek},
  {Leung}, {Masui}, {Mnchmeyer}, {Ng}, {Rafiei-Ravandi}, {Scholz}, {Shin},
  {Smith}, {Stairs}, \& {Zwaniga}}]{Bhardwaj2021}
{Bhardwaj}, M., {Gaensler}, B.~M., {Kaspi}, V.~M., {et~al.} 2021, \apjl, 910,
  L18, \dodoi{10.3847/2041-8213/abeaa6}

\bibitem[{Chen {et~al.}(2024)Chen, Jia, Dong, \& Wang}]{Chen2024}
Chen, J.~H., Jia, X.~D., Dong, X.~F., \& Wang, F.~Y. 2024, Astrophys. J. Lett.,
  973, L54, \dodoi{10.3847/2041-8213/ad7b39}

\bibitem[{Cordes \& Chatterjee(2019)}]{Cordes2019}
Cordes, J.~M., \& Chatterjee, S. 2019, Ann. Rev. Astron. Astrophys., 57, 417,
  \dodoi{10.1146/annurev-astro-091918-104501}

\bibitem[{Cordes \& Lazio(2002)}]{Cordes2002}
Cordes, J.~M., \& Lazio, T. J.~W. 2002.
\newblock \doarXiv{astro-ph/0207156}

\bibitem[{Cordes {et~al.}(2022)Cordes, Ocker, \& Chatterjee}]{Cordes2021}
Cordes, J.~M., Ocker, S.~K., \& Chatterjee, S. 2022, Astrophys. J., 931, 88,
  \dodoi{10.3847/1538-4357/ac6873}

\bibitem[{Deng \& Zhang(2014)}]{Deng2014}
Deng, W., \& Zhang, B. 2014, Astrophys. J. Lett., 783, L35,
  \dodoi{10.1088/2041-8205/783/2/L35}

\bibitem[{Foreman-Mackey {et~al.}(2013)Foreman-Mackey, Hogg, Lang, \&
  Goodman}]{emcee}
Foreman-Mackey, D., Hogg, D.~W., Lang, D., \& Goodman, J. 2013, Publ. Astron.
  Soc. Pac., 125, 306, \dodoi{10.1086/670067}

\bibitem[{Gao {et~al.}(2014)Gao, Li, \& Zhang}]{Gao2014}
Gao, H., Li, Z., \& Zhang, B. 2014, Astrophys. J., 788, 189,
  \dodoi{10.1088/0004-637X/788/2/189}

\bibitem[{Gupta {et~al.}(2025)Gupta, Beniamini, Kumar, \&
  Finkelstein}]{Gupta2025}
Gupta, O., Beniamini, P., Kumar, P., \& Finkelstein, S.~L. 2025.
\newblock \doarXiv{2501.09810}

\bibitem[{James {et~al.}(2022{\natexlab{a}})James, Prochaska, Macquart,
  North-Hickey, Bannister, \& Dunning}]{James2022a}
James, C.~W., Prochaska, J.~X., Macquart, J.~P., {et~al.} 2022{\natexlab{a}},
  Mon. Not. Roy. Astron. Soc., 509, 4775, \dodoi{10.1093/mnras/stab3051}

\bibitem[{James {et~al.}(2022{\natexlab{b}})James, Prochaska, Macquart,
  North-Hickey, Bannister, \& Dunning}]{James2022b}
---. 2022{\natexlab{b}}, Mon. Not. Roy. Astron. Soc., 510, L18,
  \dodoi{10.1093/mnrasl/slab117}

\bibitem[{Kistler {et~al.}(2008)Kistler, Yuksel, Beacom, \&
  Stanek}]{Kistler2007}
Kistler, M.~D., Yuksel, H., Beacom, J.~F., \& Stanek, K.~Z. 2008, Astrophys. J.
  Lett., 673, L119, \dodoi{10.1086/527671}

\bibitem[{Li {et~al.}(2020)Li, Gao, Wei, Yang, Zhang, \& Zhu}]{Li2020}
Li, Z., Gao, H., Wei, J.-J., {et~al.} 2020, Mon. Not. Roy. Astron. Soc., 496,
  L28, \dodoi{10.1093/mnrasl/slaa070}

\bibitem[{Li {et~al.}(2018)Li, Gao, Ding, Wang, \& Zhang}]{Li2018}
Li, Z.-X., Gao, H., Ding, X.-H., Wang, G.-J., \& Zhang, B. 2018, Nature
  Commun., 9, 3833, \dodoi{10.1038/s41467-018-06303-0}

\bibitem[{Liao {et~al.}(2020)Liao, Zhang, Li, \& Gao}]{Liao2020}
Liao, K., Zhang, S.~B., Li, Z., \& Gao, H. 2020, Astrophys. J., 896, L11,
  \dodoi{10.3847/2041-8213/ab963e}

\bibitem[{Lin {et~al.}(2024b)Lin, Li, \& Zou}]{Lin2024b}
Lin, H.-N., Li, X.-Y., \& Zou, R. 2024b, Astrophys. J., 969, 123,
  \dodoi{10.3847/1538-4357/ad5310}

\bibitem[{Lin \& Zou(2024a)}]{Lin2024a}
Lin, H.-N., \& Zou, R. 2024a, Astrophys. J., 962, 73,
  \dodoi{10.3847/1538-4357/ad1b4f}

\bibitem[{Lorimer {et~al.}(2007)Lorimer, Bailes, McLaughlin, Narkevic, \&
  Crawford}]{Lorimer2007}
Lorimer, D.~R., Bailes, M., McLaughlin, M.~A., Narkevic, D.~J., \& Crawford, F.
  2007, Science, 318, 777, \dodoi{10.1126/science.1147532}

\bibitem[{Lu {et~al.}(2020)Lu, Piro, \& Waxman}]{Lu2020}
Lu, W., Piro, A.~L., \& Waxman, E. 2020, Mon. Not. Roy. Astron. Soc., 498,
  1973, \dodoi{10.1093/mnras/staa2397}

\bibitem[{Luo {et~al.}(2020)Luo, Men, Lee, Wang, Lorimer, \& Zhang}]{Luo2020}
Luo, R., Men, Y., Lee, K., {et~al.} 2020, Mon. Not. Roy. Astron. Soc., 494,
  665, \dodoi{10.1093/mnras/staa704}

\bibitem[{Macquart {et~al.}(2019)Macquart, Shannon, Bannister, James, Ekers, \&
  Bunton}]{Macquart2019}
Macquart, J.~P., Shannon, R.~M., Bannister, K.~W., {et~al.} 2019, Astrophys. J.
  Lett., 872, L19, \dodoi{10.3847/2041-8213/ab03d6}

\bibitem[{Macquart {et~al.}(2020)}]{Macquart2020}
Macquart, J.~P., {et~al.} 2020, Nature, 581, 391,
  \dodoi{10.1038/s41586-020-2300-2}

\bibitem[{Madau \& Fragos(2017)}]{Madau2017}
Madau, P., \& Fragos, T. 2017, Astrophys. J., 840, 39,
  \dodoi{10.3847/1538-4357/aa6af9}

\bibitem[{Mandel {et~al.}(2019)Mandel, Farr, \& Gair}]{Mandel2019}
Mandel, I., Farr, W.~M., \& Gair, J.~R. 2019, Mon. Not. Roy. Astron. Soc., 486,
  1086, \dodoi{10.1093/mnras/stz896}

\bibitem[{Mastrogiovanni {et~al.}(2024)Mastrogiovanni, Pierra, Perri\`es,
  Laghi, Caneva~Santoro, Ghosh, Gray, Karathanasis, \&
  Leyde}]{Mastrogiovanni2024}
Mastrogiovanni, S., Pierra, G., Perri\`es, S., {et~al.} 2024, Astron.
  Astrophys., 682, A167, \dodoi{10.1051/0004-6361/202347007}

\bibitem[{Mu\~noz {et~al.}(2016)Mu\~noz, Kovetz, Dai, \&
  Kamionkowski}]{Munoz2016}
Mu\~noz, J.~B., Kovetz, E.~D., Dai, L., \& Kamionkowski, M. 2016, Phys. Rev.
  Lett., 117, 091301, \dodoi{10.1103/PhysRevLett.117.091301}

\bibitem[{Petroff {et~al.}(2019)Petroff, Hessels, \& Lorimer}]{Petroff2019}
Petroff, E., Hessels, J. W.~T., \& Lorimer, D.~R. 2019, Astron. Astrophys.
  Rev., 27, 4, \dodoi{10.1007/s00159-019-0116-6}

\bibitem[{Pleunis {et~al.}(2021)}]{Pleunis2021}
Pleunis, Z., {et~al.} 2021, Astrophys. J., 923, 1,
  \dodoi{10.3847/1538-4357/ac33ac}

\bibitem[{Qiang {et~al.}(2022)Qiang, Li, \& Wei}]{Qiang2022}
Qiang, D.-C., Li, S.-L., \& Wei, H. 2022, JCAP, 01, 040,
  \dodoi{10.1088/1475-7516/2022/01/040}

\bibitem[{Qiang \& Wei(2020)}]{Qiang2020}
Qiang, D.-C., \& Wei, H. 2020, JCAP, 04, 023,
  \dodoi{10.1088/1475-7516/2020/04/023}

\bibitem[{Schwarz(1978)}]{Schwarz1978}
Schwarz, G. 1978, Annals Statist., 6, 461

\bibitem[{Shin {et~al.}(2023)}]{Shin2022}
Shin, K., {et~al.} 2023, Astrophys. J., 944, 105,
  \dodoi{10.3847/1538-4357/acaf06}

\bibitem[{Tang {et~al.}(2023)Tang, Lin, \& Li}]{Tang2023}
Tang, L., Lin, H.-N., \& Li, X. 2023, Chin. Phys. C, 47, 085105,
  \dodoi{10.1088/1674-1137/acda1c}

\bibitem[{Wei {et~al.}(2015)Wei, Gao, Wu, \& M\'esz\'aros}]{Wei2015}
Wei, J.-J., Gao, H., Wu, X.-F., \& M\'esz\'aros, P. 2015, Phys. Rev. Lett.,
  115, 261101, \dodoi{10.1103/PhysRevLett.115.261101}

\bibitem[{Xiao {et~al.}(2021)Xiao, Wang, \& Dai}]{Xiao2021}
Xiao, D., Wang, F., \& Dai, Z. 2021, Sci. China Phys. Mech. Astron., 64,
  249501, \dodoi{10.1007/s11433-020-1661-7}

\bibitem[{Yao {et~al.}(2017)Yao, Manchester, \& Wang}]{Yao2017}
Yao, J.~M., Manchester, R.~N., \& Wang, N. 2017, Astrophys. J., 835, 29,
  \dodoi{10.3847/1538-4357/835/1/29}

\bibitem[{Yuksel {et~al.}(2008)Yuksel, Kistler, Beacom, \& Hopkins}]{Yukse2008}
Yuksel, H., Kistler, M.~D., Beacom, J.~F., \& Hopkins, A.~M. 2008, Astrophys.
  J. Lett., 683, L5, \dodoi{10.1086/591449}

\bibitem[{Zhang(2018)}]{Zhang2018}
Zhang, B. 2018, Astrophys. J. Lett., 867, L21, \dodoi{10.3847/2041-8213/aae8e3}

\bibitem[{Zhang(2020)}]{Zhang2020}
---. 2020, Nature, 587, 45, \dodoi{10.1038/s41586-020-2828-1}

\bibitem[{Zhang \& Wang(2019)}]{Zhang2019}
Zhang, G.~Q., \& Wang, F.~Y. 2019, Mon. Not. Roy. Astron. Soc., 487, 3672,
  \dodoi{10.1093/mnras/stz1566}

\bibitem[{Zhang {et~al.}(2023)Zhang, Zhang, Li, Yang, Cui, Ye, Wang, Yang, Bi,
  \& Zhang}]{Zhang2023}
Zhang, J., Zhang, C., Li, D., {et~al.} 2023, Astron. Rep., 67, 244,
  \dodoi{10.1134/S1063772923030083}

\bibitem[{Zhang \& Zhang(2022)}]{Zhang2022}
Zhang, R.~C., \& Zhang, B. 2022, Astrophys. J. Lett., 924, L14,
  \dodoi{10.3847/2041-8213/ac46ad}

\bibitem[{Zhang {et~al.}(2021)Zhang, Zhang, Li, \& Lorimer}]{Zhang2021}
Zhang, R.~C., Zhang, B., Li, Y., \& Lorimer, D.~R. 2021, Mon. Not. Roy. Astron.
  Soc., 501, 157, \dodoi{10.1093/mnras/staa3537}

\bibitem[{Zhou {et~al.}(2022)Zhou, Li, Liao, Niu, Gao, Huang, Huang, \&
  Zhang}]{Zhou2022}
Zhou, H., Li, Z., Liao, K., {et~al.} 2022, Astrophys. J., 928, 124,
  \dodoi{10.3847/1538-4357/ac510d}

\end{thebibliography}
\bibliographystyle{aasjournal}

\end{document}